\documentclass[12pt]{article}

\catcode`\@=11
\@addtoreset{equation}{section}

\global\arraycolsep=2pt
\usepackage{mathrsfs,amsbsy,amssymb,latexsym,amsfonts,amsmath}
\usepackage{cite}
\usepackage{graphicx,color}

\allowdisplaybreaks

\begin{document}

\begin{flushright}
\parbox{4.2cm}
{KUNS-2286}
\end{flushright}

\vspace*{2cm}

\begin{center}
{\Large \bf 
Hidden Yangian symmetry in sigma model \\ 
on squashed sphere}
\vspace*{2cm}\\
{\large Io Kawaguchi\footnote{E-mail:~io@gauge.scphys.kyoto-u.ac.jp} 
and 
Kentaroh Yoshida\footnote{E-mail:~kyoshida@gauge.scphys.kyoto-u.ac.jp} 
}
\end{center}
\vspace*{1cm}
\begin{center}
{\it Department of Physics, Kyoto University \\ 
Kyoto 606-8502, Japan} 
\end{center}

\vspace{1cm}

\begin{abstract}
We discuss a hidden symmetry of a two-dimensional sigma model on a squashed $S^3$. 
The $SU(2)$ current can be improved so that it can be regarded as a flat connection. 
Then we can obtain an infinite number of conserved non-local charges 
and show the Yangian algebra by directly checking the Serre relation. 
This symmetry is also deduced from the coset structure of the squashed sphere. 
The same argument is applicable to the warped AdS$_3$ spaces via double Wick rotations. 
\end{abstract}

\thispagestyle{empty}
\setcounter{page}{0}

\newpage

\section{Introduction}
 
It is well known that two-dimensional sigma models on {\it symmetric} 
spaces are classically integrable \cite{AAR}. Simple examples are 
the $O(3)$ non-linear sigma model and the $SU(2)$ principal chiral model. 
The integrability of the sigma models survives at quantum level and physical quantities 
such as S-matrix can be computed exactly \cite{Luscher1,Luscher2}. 
Thus the integrability allows us to study non-perturbative aspects  
of quantum field theory. There is a hidden symmetry \cite{Bernard,MacKay}, called the Yangian symmetry 
\cite{Drinfeld}, behind the integrable structure. 

The integrable structure on symmetric spaces may appear as classical integrability of 
type IIB superstring on AdS$_5\times S^5$ \cite{BPR} 
and now it has a new perspective and potential applications 
in the AdS/CFT correspondence \cite{Maldacena}. The integrable string backgrounds 
have recently been classified based on symmetric coset structure \cite{Zarembo2}.  
Then an interesting issue is to consider 
integrability and hidden symmetry of the sigma models on {\it non-symmetric} spaces 
and discuss a generalization of the AdS/CFT dictionary. Simple examples of non-symmetric spaces 
are squashed spheres and warped AdS spaces. 

In this letter, as a simple example, we will discuss a hidden symmetry of a sigma model 
on a squashed $S^3$. This squashed geometry is realized as a $U(1)$-fibration ($\psi$) 
over $S^2$ ($\theta,\phi$) and the metric is given by 
\begin{eqnarray}
ds^2 &=& \frac{L^2}{4}\left[\, d\theta^2 + \cos^2{\theta}\,d\phi^2 
+(1+C)(d\psi +\sin{\theta}\,d\phi)^2\, \right]\,. \label{sphere}
\end{eqnarray}
The constant parameter $C$ measures a deformation of $S^3$ and 
$C=0$ just describes the round $S^3$ with the radius $L$\,. 
The $SU(2)_{\rm L} \times SU(2)_{\rm R}$ symmetry for the round $S^3$ is broken to 
$SU(2)_{\rm L}\times U(1)_{\rm R}$ for $C\neq 0$\,. For $C=0$ the Yangian symmetry is well known 
as a hidden symmetry. But it remains to be clarified for $C\neq 0$  
whether or not the $SU(2)_{\rm L}\times U(1)_{\rm R}$ can enhance 
to an infinite-dimenisonal symmetry. 

\section{Yangian symmetry}

We start from the classical action of the sigma model: 
\begin{eqnarray}
S &=& -\frac{1}{2} \int\!\!\!\int\! dtdx\,
\Bigl[\,(\partial_{\mu}\theta)^2 + \cos^2{\theta}(\partial_{\mu}\phi)^2 
+(1+C)(\partial_{\mu}\psi+\sin{\theta}\partial_{\mu}\phi)^2 \Bigr]\,. \label{action}
\end{eqnarray}
For simplicity, we will not take the Virasoro conditions into account and hence 
the base space is assumed to be a two-dimensional Minkowski spacetime 
with the coordinates $x^{\mu}=(t,x)$ and the metric $\eta_{\mu\nu}=(-1,1)$\,.

The isometry of the metric (\ref{sphere}) is $SU(2)_{\rm L}\times U(1)_{\rm R}$ and hence 
the action (\ref{action}) is invariant under the following transformations:  
\begin{eqnarray}
\delta_1 \left( \phi , \psi , \theta \right) &=& \epsilon \left( -1\,,~ 0\,,~ 0 \right)\,, \nonumber \\
\delta_2 \left( \phi , \psi , \theta \right) &=& \epsilon \left( -\sin{\phi}\tan{\theta}\,,~ \frac{\sin{\phi}}{\cos{\theta}}\,,~ -\cos{\phi} \right)\,, \nonumber \\
\delta_3 \left( \phi , \psi , \theta \right) &=& \epsilon \left( \cos{\phi}\tan{\theta}\,,~ -\frac{\cos{\phi}}{\cos{\theta}}\,,~ -\sin{\phi} \right)\,, \nonumber \\
\delta_4 \left( \phi , \psi , \theta \right) &=& \epsilon \left( 0\,,~ -1\,,~ 0 \right)\,. \nonumber
\end{eqnarray}
Here $\epsilon$ is an infinitesimal constant. Note that the transformations 
are independent of the squashing parameter $C$\,. 
The first three are the $SU(2)_{\rm L}$ generators and the last is the $U(1)_{\rm R}$ one. 
By following the Noether's theorem, we can construct the four conserved currents
\begin{eqnarray}
j_{\mu}^{1} &=& \partial_{\mu}\phi +\sin{\theta}\,\partial_{\mu}\psi + C\,\sin{\theta}\,
(\partial_{\mu}\psi + \sin{\theta}\,\partial_{\mu}\phi) 
\,, \nonumber \\
j_{\mu}^{2} &=& \cos{\phi}\,\partial_{\mu}\theta - \sin{\phi}\,\cos{\theta}\,\partial_{\mu}\psi 
\nonumber \\ 
&& \qquad 
- C\sin{\phi}\,\cos{\theta}\,( \partial_{\mu}\,\psi +\sin{\theta}\,\partial_{\mu}\phi )\,, 
\nonumber \\
j_{\mu}^{3} &=& \sin{\phi}\,\partial_{\mu}\theta + \cos{\phi}\,\cos{\theta}\,\partial_{\mu}\psi 
\nonumber \\ 
&& \qquad 
+ C\,\cos{\phi}\,\cos{\theta}\,(\partial_{\mu}\psi + \sin{\theta}\,\partial_{\mu}\phi)\,, \nonumber \\
j_{\mu}^{4} &=& (1+C)(\partial_{\mu}\psi +\sin{\theta}\,\partial_{\mu}\phi )\,. \nonumber
\end{eqnarray}
After some algebra, for the $SU(2)_{\rm L}$ current $j_{\mu}^A~(A=1,2,3)$, we obtain that 
\begin{eqnarray}
\epsilon^{\mu\nu}\left(
\partial_{\mu}j_{\nu}^{A}-\frac{1}{2}
\varepsilon_{BC}^{~~~A} j_{\mu}^{B}\,j_{\nu}^{C}\right) = 
C n^A\,\epsilon^{\mu\nu}\partial_{\mu}(\sin\theta)\partial_{\nu}\phi  
\nonumber 
\end{eqnarray}
with an anti-symmetric tensor $\epsilon_{\mu\nu}$ normalized as $\epsilon_{tx}=+1$ 
and $n^A$ is a unit vector on the $S^2$
\[
n^{A} \equiv (\sin{\theta},-\sin{\phi}\cos{\theta},\cos{\phi}\cos{\theta})\,.
\]
Thus the $SU(2)_{\rm L}$ current does not satisfy the flatness condition. 
But it can be improved with the ambiguity of the Noether current 
so that it satisfies the flatness condition.  
To improve the current, the following topological current $I_{\mu}^{A}$ 
has to be added to $j_{\mu}^{A}$~:
\begin{eqnarray}
I_{\mu}^{A} \equiv \pm \sqrt{C} \epsilon_{\mu\nu}\partial^{\nu}n^{A}\,.
\label{improve} 
\end{eqnarray} 
This term is obviously conserved: $\partial^{\mu}I_{\mu}^{A}=0$\,. 
We will discuss the geometrical meaning of (\ref{improve}) later. 

By using the canonical Poisson bracket, 
the current algebra is computed as 
\begin{eqnarray}
\{ j_{t}^{A}(x),j_{t}^{B}(y) \}_{\rm P} &=& \varepsilon^{AB}_{~~~C}\,
j_{t}^{C}(x)\, \delta(x-y)\,, \nonumber  \\
\{ j_{t}^{A}(x),j_{x}^{B}(y) \}_{\rm P} &=& \varepsilon^{AB}_{~~~C}\,
j_{x}^{C}(x)\, \delta(x-y) \label{current} \\ 
&& \qquad + (1+C)\delta^{AB}\partial_{x}\delta(x-y)\,, \nonumber \\
\{ j_{x}^{A}(x),j_{x}^{B}(y) \}_{\rm P} &=& 
-C \varepsilon^{AB}_{~~~C}\,
j_{t}^{C}(x)\, \delta(x-y)\,. \nonumber
\end{eqnarray}
The algebra (\ref{current}) contains the squashing parameter $C$ and 
the last bracket $\{j_x^A,j_x^B\}_{\rm P}$ does not vanish any more. 
Hence it is not obvious whether the algebra (\ref{current}) leads to the Yangian 
algebra or not. 

After the flat conserved current has been obtained,  
an infinite number of the conserved charges can be constructed by following \cite{BIZZ}. 
The Noether charges $Q_{(0)}^{A}$ and the first non-local charges $Q_{(1)}^{A}$ 
are given by, respectively, 
\begin{eqnarray}
Q_{(0)}^{A} &\equiv& \int\!\! dx\, j_{t}^{A}(x)\,, \nonumber \\
Q_{(1)}^{A} &\equiv&  - \int\!\! dx\, j_{x}^{A}(x) + \frac{1}{4}\int\!\!\!\int\!\! dxdy\, 
\epsilon(x-y)\,\varepsilon_{BC}^{~~~A}\,
j_{t}^{B}(x)j_{t}^{C}(y)\,. \nonumber 
\end{eqnarray}
The Poisson brackets for $Q_{(0)}^A$ and $Q_{(1)}^A$ are
\begin{eqnarray}
\{ Q_{(0)}^{A},Q_{(0)}^{B} \}_{\rm P} &=& \varepsilon^{AB}_{~~~C}Q_{(0)}^{C}\,, 
\nonumber \\
\{ Q_{(0)}^{A},Q_{(1)}^{B} \}_{\rm P} &=& \varepsilon^{AB}_{~~~C}Q_{(1)}^{C}\,, 
\nonumber \\
\{ Q_{(1)}^{A},Q_{(1)}^{B} \}_{\rm P} &=& \varepsilon^{AB}_{~~~C}
\left[Q_{(2)}^{C} - \frac{1}{12}Q_{(0)}^{C}(Q_{(0)})^2 - C Q_{(0)}^{C}\right], \nonumber
\end{eqnarray}
where $Q_{(2)}^{A}$ are the second non-local charges defined as 
\begin{eqnarray}
Q_{(2)}^{A} &\equiv& \frac{1}{12}\int\!\!\!\int\!\!\!\int\!\! dxdydz\, \epsilon(x-y)\,
\epsilon(y-z)\,\delta_{BC}\, \left[\,j_{t}^{A}(x)\,j_{t}^{B}(y)\,j_{t}^{C}(z) 
- j_{t}^{B}(x)\,j_{t}^{A}(y)\,j_{t}^{C}(z)\,\right] \nonumber \\
 && \qquad + \frac{1}{2}\int\!\!\!\int\!\! dxdy\, \epsilon(x-y)\,\varepsilon_{BC}^{~~~A}\,
j_{t}^{B}(x)\,\,j_{x}^{C}(y)\,. \nonumber
\end{eqnarray}
It is straightforward to check the Serre relation:
\begin{eqnarray}
&& \{ \{ Q_{(1)}^{+},Q_{(1)}^{-} \}_{\rm P} , Q_{(1)}^{3} \}_{\rm P} = \frac{1}{4}Q_{(0)}^{3}(Q_{(1)}^{+}Q_{(0)}^{-}-Q_{(1)}^{-}Q_{(0)}^{+})\,, 
\nonumber \\
&& \{ \{ Q_{(1)}^{3},Q_{(1)}^{\pm} \}_{\rm P} , Q_{(1)}^{\pm} \}_{\rm P} 
= \frac{1}{4}Q_{(0)}^{\pm}(Q_{(1)}^{3}Q_{(0)}^{\pm}-Q_{(1)}^{\pm}Q_{(0)}^{3})\,, \nonumber \\
&& \{ \{ Q_{(1)}^{+},Q_{(1)}^{-} \}_{\rm P} , Q_{(1)}^{\pm} \}_{\rm P} \pm 2 \{ \{ Q_{(1)}^{3},Q_{(1)}^{\pm} \}_{\rm P} , Q_{(1)}^{3} \}_{\rm P} \nonumber \\
&=& \frac{1}{4}Q_{(0)}^{\pm}(Q_{(1)}^{+}Q_{(0)}^{-}-Q_{(1)}^{-}Q_{(0)}^{+}) 
\pm \frac{1}{2}Q_{(0)}^{3}(Q_{(1)}^{3}Q_{(0)}^{\pm}-Q_{(1)}^{\pm}Q_{(0)}^{3})\,. \nonumber
\end{eqnarray}
Thus we have shown that the non-local charges constructed from the improved flat current 
surely satisfy the Yangian algebra. 

This result may seem curious. The Yangian algebra should correspond 
to XXX model, while the squashed sphere is described 
as a deformation of the round $S^3$ and the resulting algebra is expected to be 
an affine quantum algebra related to XXZ model rather than XXX model.  
In fact, the discretized model, which gives rise to the sigma model on the squashed 
$S^3$ in a continuum limit, is XXZ model as discussed in \cite{FR}. 

\begin{figure}
\begin{center}
\includegraphics[scale=.4]{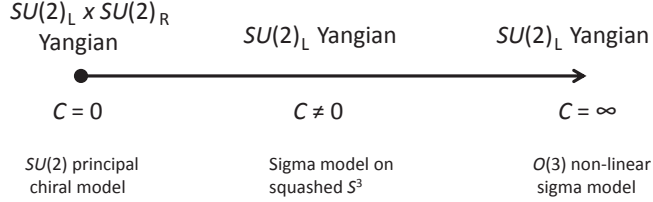}%
\caption{\footnotesize Yangians with respect to the squashing parameter $C$.
\label{Yangians}}
\end{center}
\end{figure}

How can one understand the Yangian algebra? To find a natural interpretation, 
it would be helpful to consider the $C\to\infty$ limit, 
where the geometry is reduced to the direct product $S^2\times S^1$. Then 
it is easy to find the $SU(2)_{\rm L}$ Yangian from the $O(3)$ non-linear sigma model.  
On the other hand, for $C=0$ there is the $SU(2)_{\rm L} \times SU(2)_{\rm R}$ 
Yangian. A natural interpretation is that the Yangian obtained above is inherited from  
the $SU(2)_{\rm L}$ Yangian for $C=\infty$ to for finite $C$ 
and interpolate to the $SU(2)_{\rm L}$ Yangian subalgebra for $C=0$, 
as depicted in Fig.\ \ref{Yangians}. In other words, the $SU(2)_{\rm L}$ Yangian symmetry 
of the $O(3)$ non-linear sigma model can survive under the Hopf $U(1)$ fibration for all values 
of $C$. 

An interesting question here is whether the Yangian 
symmetry means the complete integrability of the sigma model on the squashed sphere 
in the sense of Liouville. There are an infinite number of conserved charges,  
but which would not imply the complete integrability. 
Even if the Yangian symmetry does not mean the complete 
integrability, we may expect some strong constraints coming 
from the infinite-dimensional symmetry and they may be useful 
to compute some physical quantities, for example, in the the context of AdS/CFT. 

Furthermore it is worth noting the physical interpretation of the improvement term. 
The topological current (\ref{improve}) can directly be obtained 
if the Hopf term is added to the classical action 
(\ref{action})~:
\begin{eqnarray}
S_{\rm Hopf} &=& \pm \sqrt{C}\int\!\!\!\int\!\! dtdx\,\epsilon^{\mu\nu}\,\cos{\theta}\,
\partial_{\mu}\theta\,\partial_{\nu}\phi \nonumber \\ 
&=&\pm\frac{\sqrt{C}}{2}\int\!\!\!\int\!\! dtdx\,\epsilon^{\mu\nu}\,
\varepsilon_{ABC}\,n^A\partial_{\mu}n^B\partial_{\nu}n^C
\,. \nonumber 
\end{eqnarray} 
But this term vanishes automatically because there is no topological obstruction 
in the present setup, namely $\pi_2(S^3)=0$, and it can be gauged away. 
Thus it does not change the original theory but it is still sensitive to 
the flatness condition via the ambiguity of the Noether current.  

A similar geometry, that is the squashed $S^3$ geometry equipped with the two-form flux,  
can be realized as a squashed giant graviton \cite{giant} 
and hence the hidden symmetry found here may be useful in this issue. 
For example, it would be possible to consider an open string attaching on the squashed 
giant graviton by generalizing the works on the maximal giant graviton case \cite{BV, 
Agarwal,OY,HM}. This would be an interesting direction to be studied. 

\section{Coset construction of squashed $S^3$}

It is quite remarkable that the Yangian symmetry has been found in the sigma model 
on the squashed $S^3$, because the squashed $S^3$ is NOT described as a symmetric coset. 
Thus it should be interesting to see the coset structure of the squashed $S^3$. 

The squashed $S^3$ has the isometry $G=SU(2)_{\rm L}\times U(1)_{\rm R}$\,. 
The $SU(2)_{\rm L}$ generators $T_{A}$ satisfy
 $[T_{A},T_{B}]=\varepsilon_{AB}^{~~~~C} T_{C}$\,. 
The $U(1)_{\rm R}$ generator is represented by $T_4$\,. 

Let us consider a coset $M = G/H$\,. When the Lie algebras for $M$ and $H$ are written as  
$\mathfrak{m}$ and $\mathfrak{h}$\,, respectively, we may consider the following algebras
\begin{eqnarray}
\mathfrak{m}=\left\{ T_{1},T_{2},T_{3} \right\}\,, \qquad 
\mathfrak{h}=\{ \alpha T_{3} + \beta T_4 \}\,. \nonumber 
\end{eqnarray}
Here $\alpha$ and $\beta$ are constant parameters. 
The group structure constant related to our argument is represented by 
\begin{eqnarray}
f_{[m]\alpha T_{3}+\beta T_4}^{~~~~~~~~~~~~~ [n]} = \begin{pmatrix}
0 & -\alpha & 0 \\ 
\alpha & 0 & 0 \\ 
0 & 0 & 0
\end{pmatrix}
\,, \nonumber 
\end{eqnarray}
where the indices $[m],[n]$ are for $\mathfrak{m}$ and are defined up to 
$\mathfrak{h}$-transformation. 

The most general symmetric two-form  is given by \cite{NW}
\begin{eqnarray}
\Omega_{[m][n]} = \begin{pmatrix}
\Omega_{11} & 0 & 0 \\ 
0 & \Omega_{11} & 0 \\ 
0 & 0 & \Omega_{33}
\end{pmatrix}
\,. \nonumber 
\end{eqnarray}
By taking a coset representative as $g = {\rm e}^{\phi T_{1}} {\rm e}^{\theta T_{2}} 
{\rm e}^{\psi T_{3}}$ 
and expanding the Maurer-Cartan one-form $J=g^{-1}dg$\,, the vielbeine are computed as 
\begin{eqnarray}
e^{1} &=& \cos\psi\,\cos\theta\,d\phi + \sin\psi\,d\theta\,, \nonumber \\ 
e^{2} &=& -\sin\psi\,\cos\theta\,d\phi + \cos\psi\,d\theta\,, \nonumber \\ 
e^{3} &=& d\psi + \sin\theta\,d\phi\,. \nonumber
\end{eqnarray}
Taking $\Omega_{11}=L^2/4$ and $\Omega_{33}=L^2(1+C)/4$, 
we obtain 
\begin{eqnarray}
ds^2 &=& \Omega_{[m][n]}e^{[m]}e^{[n]} \nonumber \\ 
&=& \frac{L^2}{4}
\left[
\cos^2{\theta}d\phi^2
+d\theta^2
+(1+C)(d\psi+\sin{\theta}\,d\phi)^2 
\right]
\,. \nonumber 
\end{eqnarray}
Thus the metric of the squashed $S^3$ has been reproduced 
by coset construction. Note that the presense of $U(1)_{\rm R}$ generator $T_4$ 
is crucial for the construction. If not, the coset cannot be defined. 

Let us comment on the relation between the coset structure and non-local charges. 
The coset we used here is not symmetric but satisfies 
$[\mathfrak{m},\mathfrak{m}]\subset\mathfrak{m}$\,. For this case as well as symmetric cosets,   
an infinite set of non-local charges can be constructed potentially as described in \cite{AAR}. 
Thus the coset structure would also be consistent to the previous result. 

\section{From squashed $S^3$ to warped AdS$_3$}

We can discuss warped AdS$_3$ spaces in the same way as in the squashed $S^3$ case.  
As summarized in \cite{warped}, there are three kinds of warped AdS$_3$ spaces: 
1) space-like, 2) time-like and 3) null. The metrics of the first two are reproduced from 
the metric of the squashed $S^3$ (\ref{sphere}) via double Wick rotations as follows.  

First, by taking a Wick rotation 
$\theta \rightarrow i\sigma\,,~\phi \rightarrow \tau\,,~ 
\psi \rightarrow iu$ 
and inverting the overall sign, we obtain the metric of space-like warped AdS$_3$ 
\begin{eqnarray}
ds^2 &=& \frac{L^2}{4}\left[ -\cosh^2{\sigma}\,d\tau^2 +d\sigma^2 
+(1+C)(du +\sinh{\sigma}\,d\tau)^2 \right]\,. \nonumber 
\end{eqnarray}
Similarly, by taking another Wick rotation 
$\theta \rightarrow i\sigma\,,~\phi \rightarrow iu\,,~\psi \rightarrow \tau$ 
and inverting the overall sign, the metric of time-like warped AdS$_3$ is obtained as  
\begin{eqnarray}
ds^2 &=& \frac{L^2}{4}\left[ \cosh^2{\sigma}\,du^2 +d\sigma^2 
-(1+C)(d\tau -\sinh{\sigma}\,du)^2 \right]\,. \nonumber 
\end{eqnarray}
The hidden symmetry of sigma models on space-like and time-like warped AdS$_3$ spaces 
can be shown in the same way. The coset structure is also the same. 

For the null case 3), the metric is known as 
a three-dimensional Schr$\ddot{\rm o}$dinger spacetime 
\cite{Son,BM} and not directly related to the squashed $S^3$\,, 
though the metric can be obtained by coset construction \cite{SYY}. 
Then the coset proposed in \cite{SYY} is non-reductive and the hidden symmetry  
of sigma models on the null warped AdS$_3$ remains to be clarified. 

It is known that the metrics of wapred AdS spaces become solution of topologically massive 
gravity \cite{TMG} by taking an appropriate value of $C$ (depending on the gravitational Chern-Simons coupling and cosmological constant). The hidden symmetry may play an important role 
in studying topologically massive gravity. 

\section{Discussions}

There are many applications. 
First, the hidden symmetry of warped AdS$_3$ spaces and squashed $S^3$ may be useful 
in considering a deformation of spin chains in AdS$_3$/CFT$_2$ discussed in \cite{Zarembo1} 
(See \cite{Wen} for an earlier attempt in this direction). 
In fact, warped AdS$_3$ and squashed $S^3$ geometries are realized as string backgrounds 
\cite{S1,S2,S3,OU,S4}. 
It would be a good exercise to show the relation 
between the Yangian symmetry and T-duality utilized in \cite{OU}. 

An easy task is to incorporate supersymmetries, for example, 
based on $SU(2|1)$ and find out a hidden super Yangian symmetry. 
It is also nice to consider a generalization to higher dimensional cases. 
A simple derivation of the metrics of squashed spheres is discussed in \cite{Hatsuda} and 
the result therein would be helpful for this purpose. 

A challenging issue is to check whether or not 
the similar argument is applicable to the integrable structure in the Kerr/CFT correspondence 
\cite{Kerr/CFT}, where a three-dimensional slice of the near-horizon extreme Kerr (NHEK) 
geometry \cite{BH} is described as a warped AdS$_3$ space. 
The sigma model on the NHEK geometry might possess a hidden symmetry 
and if it exists, it would be very helpful to elaborate 
the Kerr/CFT correspondence. 

As another issue, a warped AdS geometry 
appears also in an application of AdS/CFT to condensed matter physics \cite{Kraus}.  
The Yangian symmetry may be useful in this direction.  

The squashed geometries often appear in various ways and in many cases. 
We believe that the hidden symmetry of the sigma models 
on the geometries 
would be an important key ingredient to consider some generalizations 
of AdS/CFT correspondence.

\subsection*{Acknowledgments}
We would like to thank S.~Detournay, M.~Hatsuda, Y.~Hatsuda, K.~Iwasaki, H.~Kawai, 
D.~Orlando, R.~Sasaki, M.~Sakaguchi, K.~Sakai, Y.~Yasui and C.~A.~S.~Young for illuminating discussions. 
This work was supported by the scientific grants from the ministry of education, science,
sports, and culture of Japan (No.\,22740160), and in part by 
the Grant-in-Aid for the Global COE Program ``The Next Generation of Physics, Spun 
from Universality and Emergence'' from the Ministry of Education, Culture, Sports, 
Science and Technology (MEXT) of Japan. 
In addition, we thank the Yukawa Institute for Theoretical Physics at Kyoto University. 
Discussions during the YITP workshop YITP-W-08-04 on ``Development of Quantum Field Theory 
and String Theory'' were useful to complete this work.


\begin{thebibliography}{99}

\bibitem{AAR} 
 E.~Abdalla, M.~C.~Abdalla and K.~Rothe, 
 ``Non-perturbative methods in two-dimensional quantum field theory,'' 
 World Scientific, 1991. 

\bibitem{Luscher1}
  M.~Luscher,
  ``Quantum Nonlocal Charges And Absence Of Particle Production In The
  Two-Dimensional Nonlinear Sigma Model,''
  Nucl.\ Phys.\  B {\bf 135} (1978) 1.
  
\bibitem{Luscher2}
  M.~Luscher and K.~Pohlmeyer,
  ``Scattering Of Massless Lumps And Nonlocal Charges In The Two-Dimensional
  Classical Nonlinear Sigma Model,''
  Nucl.\ Phys.\  B {\bf 137} (1978) 46. 
  
\bibitem{Bernard}
 D.~Bernard,
gHidden Yangians In 2-D Massive Current Algebras,h 
 Commun.\ Math.\ Phys.\ {\bf 137} (1991) 191.

\bibitem{MacKay}  
N.~J.~MacKay, gOn the classical origins of Yangian symmetry in integrable field theory,h
Phys.\ Lett.\ B {\bf 281} (1992) 90 [Erratum-ibid.\ B {\bf 308} (1993) 444].
  
\bibitem{Drinfeld}
  V.~G.~Drinfel'd,
  ``Hopf Algebras and the Quantum Yang-Baxter Equation,'' 
Sov.\ Math.\ Dokl.\ {\bf 32} (1985) 254; 
  ``A New realization of Yangians and quantized affine algebras,''
  Sov.\ Math.\ Dokl.\  {\bf 36} (1988) 212.
  
\bibitem{BPR}
  I.~Bena, J.~Polchinski and R.~Roiban,
  ``Hidden symmetries of the AdS$_5\times S^5$ superstring,''
  Phys.\ Rev.\  D {\bf 69} (2004) 046002
  [arXiv:hep-th/0305116].


\bibitem{Maldacena}  
  J.~M.~Maldacena,
  ``The large N limit of superconformal field theories and supergravity,''
  Adv.\ Theor.\ Math.\ Phys.\  {\bf 2} (1998) 231
  [Int.\ J.\ Theor.\ Phys.\  {\bf 38} (1999) 1113]
  [arXiv:hep-th/9711200].

\bibitem{Zarembo2}
  K.~Zarembo,
  ``Strings on Semisymmetric Superspaces,''
  arXiv:1003.0465 [hep-th].

  
\bibitem{BIZZ}  
  E.~Brezin, C.~Itzykson, J.~Zinn-Justin and J.~B.~Zuber,
  ``Remarks About The Existence Of Nonlocal Charges In Two-Dimensional
  Models,''
  Phys.\ Lett.\  B {\bf 82} (1979) 442.  
  
\bibitem{FR}
  L.~D.~Faddeev and N.~Y.~Reshetikhin,
  ``Integrability Of The Principal Chiral Field Model In (1+1)-Dimension,''
  Annals Phys.\  {\bf 167} (1986) 227.  

\bibitem{giant}
  S.~Prokushkin and M.~M.~Sheikh-Jabbari,
  ``Squashed giants: Bound states of giant gravitons,''
  JHEP {\bf 0407} (2004) 077
  [arXiv:hep-th/0406053]. \\ 
  M.~Ali-Akbari and M.~M.~Sheikh-Jabbari,
  ``Electrified BPS Giants: BPS configurations on Giant Gravitons with Static
  Electric Field,''
  JHEP {\bf 0710} (2007) 043
  [arXiv:0708.2058 [hep-th]].  

\bibitem{BV}
  D.~Berenstein and S.~E.~Vazquez,
  ``Integrable open spin chains from giant gravitons,''
  JHEP {\bf 0506} (2005) 059
  [arXiv:hep-th/0501078].
  
\bibitem{Agarwal}
  A.~Agarwal,
  ``Open spin chains in super Yang-Mills at higher loops: Some potential
  problems with integrability,''
  JHEP {\bf 0608} (2006) 027 \\{} 
  [arXiv:hep-th/0603067].  

\bibitem{OY}
  K.~Okamura and K.~Yoshida,
  ``Higher loop Bethe ansatz for open spin-chains in AdS/CFT,''
  JHEP {\bf 0609} (2006) 081
  [arXiv:hep-th/0604100].

\bibitem{HM}
  D.~M.~Hofman and J.~M.~Maldacena,
  ``Reflecting magnons,''
  JHEP {\bf 0711} (2007) 063
  [arXiv:0708.2272 [hep-th]].
  
\bibitem{NW}
  C.~R.~Nappi and E.~Witten,
  ``A WZW model based on a nonsemisimple group,''
  Phys.\ Rev.\ Lett.\  {\bf 71} (1993) 3751
  [arXiv:hep-th/9310112].


\bibitem{warped} 
  D.~Anninos, W.~Li, M.~Padi, W.~Song and A.~Strominger,
  ``Warped AdS$_3$ Black Holes,''
  JHEP {\bf 0903} (2009) 130
  [arXiv:0807.3040 [hep-th]].
  

\bibitem{Son}
  D.~T.~Son,
  ``Toward an AdS/cold atoms correspondence: a geometric realization of the
  Schroedinger symmetry,''
  Phys.\ Rev.\  D {\bf 78} (2008) 046003 [arXiv:0804.3972 [hep-th]]. 
\bibitem{BM}  
  K.~Balasubramanian and J.~McGreevy,
  ``Gravity duals for non-relativistic CFTs,''
  Phys.\ Rev.\ Lett.\  {\bf 101} (2008) 061601
  [arXiv:0804.4053 [hep-th]].

\bibitem{SYY}
  S.~Schafer-Nameki, M.~Yamazaki and K.~Yoshida,
  ``Coset Construction for Duals of Non-relativistic CFTs,''
  JHEP {\bf 0905} (2009) 038
  [arXiv:0903.4245 [hep-th]].


\bibitem{TMG}
  S.~Deser, R.~Jackiw and S.~Templeton,
  ``Three-Dimensional Massive Gauge Theories,''
  Phys.\ Rev.\ Lett.\  {\bf 48} (1982) 975; 
  S.~Deser, R.~Jackiw and S.~Templeton,
  ``Topologically massive gauge theories,''
  Annals Phys.\  {\bf 140} (1982) 372
  [Erratum-ibid.\  {\bf 185} (1988\ APNYA,281,409-449.2000) 406.1988\ APNYA,281,409].

\bibitem{Zarembo1}
  A.~Babichenko, B.~Stefanski and K.~Zarembo, 
  ``Integrability and the AdS$_3$/CFT$_2$ correspondence,''
  arXiv:0912.1723 [hep-th].

\bibitem{Wen}
  W.~Y.~Wen,
  ``Spin chain from marginally deformed AdS$_3\times S^3$,''
  Phys.\ Rev.\  D {\bf 75} (2007) 067901
  [arXiv:hep-th/0610147]. 
  
\bibitem{S1}
  D.~Israel, C.~Kounnas, D.~Orlando and P.~M.~Petropoulos,
  ``Electric / magnetic deformations of $S^3$ and AdS$_3$, and geometric cosets,''
  Fortsch.\ Phys.\  {\bf 53} (2005) 73
  [arXiv:hep-th/0405213]; 
``Heterotic strings on homogeneous spaces,''
  Fortsch.\ Phys.\  {\bf 53} (2005) 1030
  [arXiv:hep-th/0412220].

\bibitem{S2}
  S.~Detournay, D.~Orlando, P.~M.~Petropoulos and P.~Spindel,
  ``Three-dimensional black holes from deformed anti de Sitter,''
  JHEP {\bf 0507} (2005) 072
  [arXiv:hep-th/0504231].

\bibitem{S3} 
  G.~Compere, S.~Detournay and M.~Romo,
  ``Supersymmetric G\'odel and warped black holes in string theory,''
  Phys.\ Rev.\  D {\bf 78} (2008) 104030
  [arXiv:0808.1912 [hep-th]].
 
   
\bibitem{OU}
  D.~Orlando and L.~I.~Uruchurtu,
  ``Warped anti-de Sitter spaces from brane intersections in type II string
  theory,''
  arXiv:1003.0712 [hep-th].

\bibitem{S4}
  S.~Detournay, D.~Israel, J.~M.~Lapan and M.~Romo,
  ``String Theory on Warped AdS$_3$ and Virasoro Resonances,''
  arXiv:1007.2781 [hep-th].
  
\bibitem{Hatsuda}
  M.~Hatsuda and S.~Tomizawa,
  ``Coset for Hopf fibration and Squashing,''
  Class.\ Quant.\ Grav.\  {\bf 26} (2009) 225007
  [arXiv:0906.1025 [hep-th]].

\bibitem{Kerr/CFT}  
  M.~Guica, T.~Hartman, W.~Song and A.~Strominger,
  ``The Kerr/CFT Correspondence,''
  Phys.\ Rev.\  D {\bf 80} (2009) 124008
  [arXiv:0809.4266 [hep-th]].  
  
\bibitem{BH}
  J.~M.~Bardeen and G.~T.~Horowitz,
  ``The extreme Kerr throat geometry: A vacuum analog of AdS$_2\times$S$^2$,''
  Phys.\ Rev.\  D {\bf 60} (1999) 104030
  [arXiv:hep-th/9905099].
  
\bibitem{Kraus}
 E.~D'Hoker and P.~Kraus,
  ``Charged Magnetic Brane Solutions in AdS$_5$ and the fate of the third law of
  thermodynamics,''
  JHEP {\bf 1003} (2010) 095
  [arXiv:0911.4518 [hep-th]]; 
  ``Holographic Metamagnetism, Quantum Criticality, and Crossover Behavior,''
  arXiv:1003.1302 [hep-th].


\end{thebibliography}
\end{document}